\documentclass[aip,preprint]{revtex4-1}

\usepackage{graphicx,color,bm,natbib, times, placeins} 
\usepackage{hyperref}
\usepackage{mathtools}%
\usepackage[utf8]{inputenc}
\usepackage{natbib}
\usepackage{graphicx}
\usepackage{dcolumn}
\usepackage{bm}
\usepackage{amsmath}
\usepackage{booktabs}

\draft 

\begin{document}
\title{The relationship between sentiment score and COVID-19 cases in the United States}

\author{Truong (Jack) Luu}
\affiliation{School of Information Technology, Illinois State University, Normal, IL, USA}
\author{Rosangela Follmann}
\email{rfollma@ilstu.edu}
\thanks{Accepted for publication at Journal of Information Science. \href{https://doi.org/10.1177/01655515211068167}[https://doi.org/10.1177/01655515211068167]}
\affiliation{School of Information Technology, Illinois State University, Normal, IL, USA}

\date{\today}

\begin{abstract}
The coronavirus disease (COVID-19) continues to have devastating effects across the globe. No nation has been free from the uncertainty brought by this pandemic. The health, social and economic tolls associated with it are causing strong emotions and spreading fear in people of all ages, genders, and races. Since the beginning of the COVID-19 pandemic, many have expressed their feelings and opinions related to a wide range of aspects of their lives via Twitter. In this study, we consider a framework for extracting sentiment scores and opinions from COVID-19 related tweets. We connect users' sentiment with COVID-19 cases across the USA and investigate the effect of specific COVID-19 milestones on public sentiment. The results of this work may help with the development of pandemic-related legislation, serve as a guide for scientific work, as well as inform and educate the public on core issues related to the pandemic.

\end{abstract}


\maketitle

\begin{quotation}
\end{quotation}

\section{Introduction}
Coronavirus disease (COVID-19) is an infectious disease caused by a novel coronavirus \cite{Coronavi29:online}. The first confirmed case of COVID‑19 was reported in early December 2019 in Wuhan City, Hubei Province, China. On March 11, 2020, the COVID-19 outbreak was officially declared a pandemic by the World Health Organization (WHO) \cite{WHODirec92:online}.  As of June 2021, more than 177 million cases have been reported across 220 countries and territories, resulting in more than 3.8 million deaths worldwide. The United States alone accounts for about 600 thousand deaths \cite{WHOCoron61:online,Countrie5:online}.

The high mortality and transmission rates of COVID-19, along with the consequent social-economic impacts, have brought fear and stress across the globe. According to the United Nations, COVID-19 can push more than 34 million people into extreme poverty by 2020. The global economy is predicted to lose almost \$8.5 trillion in output by 2022, cancelling out nearly all the global social-economical gain of the last four years \cite{COVID19t32:online}. In the United States, the unemployment rate averaged 5.76\% from 1948 to early 2020. During COVID-19, it reached an all-time high of 14.70\% in April of 2020 \cite{Unemploy3:online}. No nation has been free from the uncertainty brought by the pandemic and is causing strong emotions and spreading fear in all peoples of all ages, gender, and race. 

As the world we are living in today becomes even more connected throughout, information is being easily accessed and shared via various online outlets. People are, knowingly or not, giving away opinions, feelings, and personal data.
One popular form of sharing information is through microblogging which is a form of communication on the web where brief text messages, referred to as tweets, are broadcast to the public. Twitter users usually tweet about themselves or share the news. In either case, the tweets usually convey information about the mood of their authors \cite{Naaman2010}. 

The tweeted texts can be studied using empirical analysis which provides quantitative measures to allow a better understanding of the collection of texts, as for example subjectivity and polarity scores. Subjectivity reflects the amount of personal opinion and factual information contained in the text. Polarity reflects the emotions expressed in a text as positive (a good feeling) or negative (not a good feeling).  
Extracting sentiments from texts is part of the field of natural language processing and is referred to as sentiment analysis. Sentiment analysis methods were originally based on a dictionary \cite{Miller1990WordNet}, and have evolved to include machine learning algorithms \cite{Pang2002}.
These analysis tools can help in creating new health policies and in businesses' decision-making processes \cite{Ji2015,morente2019carrying}. In particular, microblogging content has an important role as a source of data for tracking disease outbreaks and helping understand public attitudes and behaviours during a crisis \cite{al2016using}. There has been much work on sentiment related to COVID-19 \cite{barkur2020sentiment,samuel2020covid,Chakraborty2020,Sarker2020}, but not much is known about the relationship between the sentiments and the COVID-19 both the number of confirmed cases and the death toll in the USA. 

In this study, we analyze COVID-19 related tweets that were generated in the USA from March 19 to August 31, 2020. We test the correlation between users' sentiment and COVID-19 cases across the USA and investigate the effect of specific COVID-19 milestones on sentiment scores. Our implementation of sentiment analysis shows the existence of a link between sentiment scores, COVID-19 confirmed cases and death toll. Additionally, significant events such as new regulations from the government, celebration of important holidays and social conflicts seem to directly affect the public's sentiment.  

\section{Related Work}

The COVID-19 pandemic has challenged our way of living. It has limited our social interactions, prompted vast virtualization of our daily routines, and promoted extensive transformations in the workplace. Such drastic changes have affected human behaviour and are having a great impact on people's mental health. Due in part to the large restrictions to in-person social interactions imposed by the COVID pandemic, about 42\% of adults in the USA  have reported symptoms of anxiety or depression in December 2020 compared to 11\% in the previous year \cite{Czeisler2020,Abbott2021}. 
Social interaction plays a crucial role in people's manifestations of emotions and sentiments by using corporal or oral expressions, or in writing. 
Confronted with severe limitations in their in-person communication capability, people resorted more heavily to social media as a means for expressing emotions and sentiments. In particular, Twitter became very popular as a written form of microblogging. It has more than 150 million users where people gather news information as well as express their concerns, feelings, and health-related information.
This extensive availability of social media data has propitiated much research work in the field of sentiment analysis.

Sentiment analysis is a field of study that uses natural language processing techniques to extract opinions and feelings from written texts and has been incorporated in areas like business, economics, and health \cite{morente2019carrying,yu2013impact,Bollen2011,Ji2015}. 
There are different approaches to sentiment analysis that can be lexicon-based \cite{Miller1990WordNet,Cambria2017,Bonta2019Lexicon} or machine learning-based \cite{Pang2002,Pang2008,Agarwal2011,Li2020user}. 
A sentiment lexicon with words and phrases predefined as positive or negative is used in the lexicon approach. In the machine learning approach training data is required for automatically classifying the text. 
Morente-Molinera et al. \cite{morente2019carrying} used lexicon-based sentiment analysis to extract preferences and build a decision-making process, while Ji et al. \cite{Ji2015} developed a two-step approach combining a corpus of personal clues and machine learning to classify Twitter sentiment for addressing public health concerns. 
Besides providing insights about people’s emotions and feelings, sentiment analysis along with text mining can provide much help in creating systemic reviews of literature related to infectious diseases \cite{Alamoodi2021,Cheng2020}. For example, studies of this kind can help health and medical communities to extract useful information and interrelationships from coronavirus-related studies, along with future directions of research topics \cite{Bose2021}.

The COVID-19 pandemic has prompted various studies trying to identify human emotional responses and opinions to the pandemic across the globe \cite{samuel2020covid,barkur2020sentiment,Chakraborty2020,Zhu2020,LasHeras-Pedrosa2020,Pokharel2020,Prabhakar2020,Imran2020,Garcia2021}.  In the early stages of the pandemic, Han et al. \cite{Han2020} explored COVID-19 related public opinion in China from January 9 to February 10, 2020. The authors used the latent Dirichlet allocation model for topic extraction, suggesting a temporal variability of the number of texts for different topics and subtopics corresponding to the different developmental stages of the event.
By looking at temporal changes and spatial distribution of COVID-19 related texts, they found a synchronization between frequent daily discussions and the trends in the COVID-19 outbreak.
Also early in the pandemic, Samuel et al. \cite{samuel2020covid} studied issues in public sentiment in the United States, reflecting concerns about Coronavirus with growth in fear and negative sentiments.  The authors used exploratory and descriptive textual analytics, along with textual data
visualization, to provide insights into the progress of fear sentiment over time as COVID-19 approached peak levels. Additionally, their work contributes to the strategic process, presenting methods with valuable informational and public sentiment insights, which can be used to develop much needed motivational solutions and strategies to counter the rapid spread of  fear-panic-despair associated with Coronavirus and COVID-19.

Li et al. \cite{Li2020impact} explored the impact of COVID-19 on people’s mental health. Texts from Weibo active users along with machine-learning predictive models, were used to compute word frequency, and scores of emotional and cognitive indicators before and after the declaration of COVID-19 on 20 January, 2020. Their results
showed that negative emotions (e.g., anxiety, depression and indignation) and sensitivity to social
risks increased, while the scores of positive emotions (e.g., Oxford happiness) and life satisfaction
decreased, suggesting a need for clinical practitioners prepare to deliver corresponding therapy foundations for the risk groups and affected people. In addition, Sarker et al. \cite{Sarker2020} showed that self-reported COVID-19 symptoms by Twitter users can complement those identified in clinical settings.
Barkur \cite{barkur2020sentiment} used sentiment analysis of tweets from India after the announcement of the lockdown, addressing the population feelings towards the lockdown, 
 while de Las Heras -- Pedrosa et al. \cite{LasHeras-Pedrosa2020} addressed the question of how social media has affected risk communication in uncertain contexts, and its impact on the emotions and sentiments derived from the semantic analysis in Spanish society during the COVID-19 pandemic. Similarly,
Chakraborty et al. \cite{Chakraborty2020} analysed COVID-19 related tweets and retweets showing how popularity is affecting accuracy in social media. 
Using topical analysis Garcia \& Berton \cite{Garcia2021} showed similar sentiment trends in the discourse evolution of COVID-19 tweets in Brazil (Portuguese language) and in the USA (English language).

In all, it is not surprising that the COVID-19 pandemic is having a devastating effect economically and emotionally across the globe. In this study, we consider a framework for extracting sentiment scores and opinions from COVID-19 related tweets in the USA and investigate the effect of COVID-19 milestones on people's sentiment.

\section{Methodology}
\begin{figure}[h]  
\centering
\includegraphics[width=0.55\textwidth]{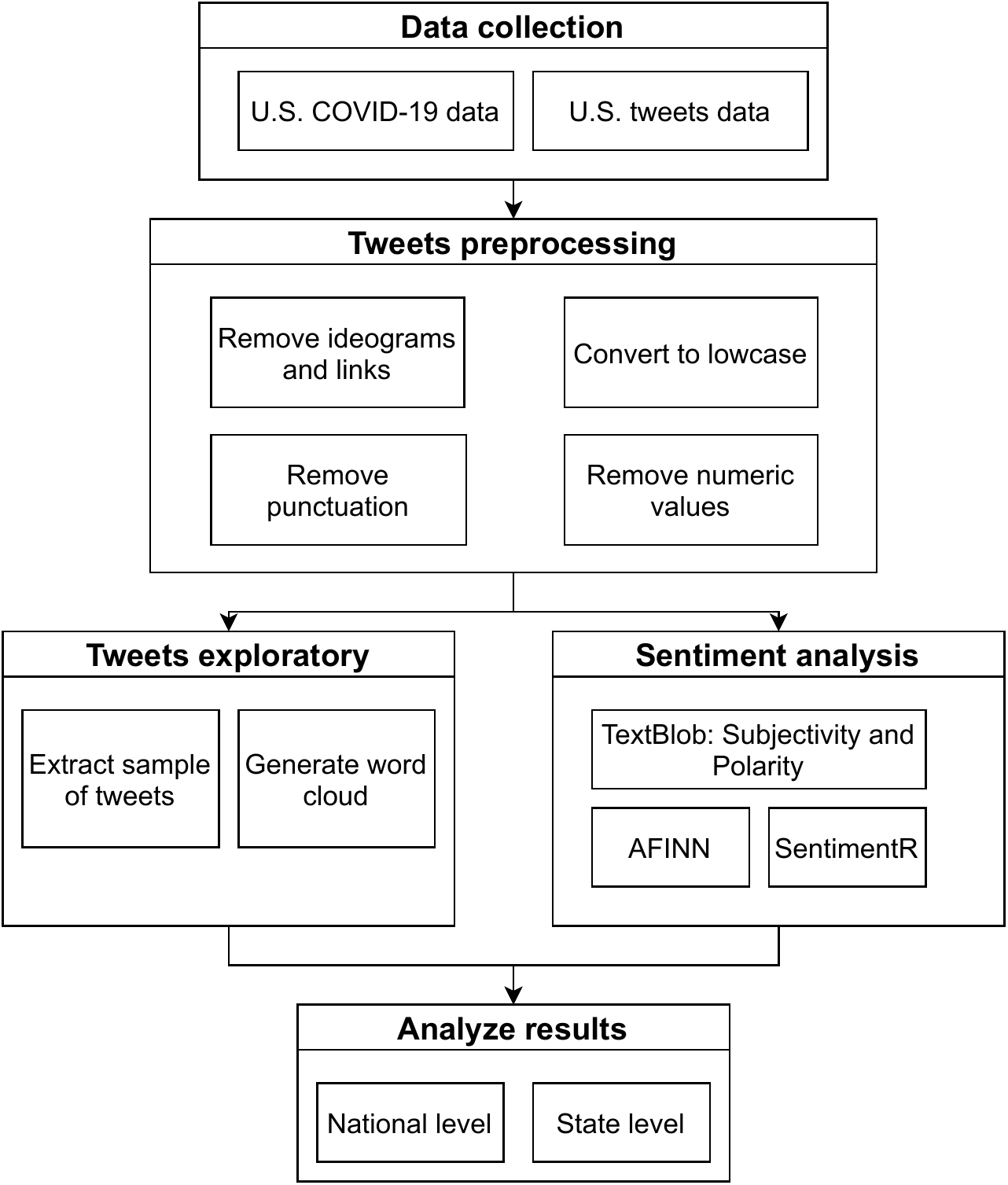}
\caption{Schematic view of the methodology used for tweet data gathering and processing procedure.} 
\label{fig1}
\end{figure}

In this study, we utilize sentiment analysis to identify outputs and trends in attitudes, feelings and opinions based on tweets in the USA during the COVID pandemic. Our approach includes collecting the COVID-19 related Tweeter data as well as the corresponding COVID-19 cases and death toll numbers. The Tweeter data are then preprocessed, and sentiment analysis is performed using three sentiment lexicons (TextBlob, AFINN and SentimentR). Next, we analyze and report the results.
A schematic view of the methodology framework is illustrated in Figure~\ref{fig1}. 

\subsection{Data Collection}
\subsubsection{Tweeter dataset.} The data we use in this work were obtained from a collection of geotagged tweet identifiers related to the COVID-19 pandemic \cite{Lamsal2020}. They included only tweets that had keywords associated to COVID-19, such as \#corona, \#coronavirus, \#covid, \#covid19, \#covid-19, and \#sarscov2, for example. Our dataset contains a total of 85,085 tweets obtained from March 19, 2020, to August 31, 2020. Each tweet identifier was hydrated by providing the text content, date, time, hashtags, and coordinates, among other attributes. The tweets were selected in association with unique geographical coordinates, with geotagged enabled by the user and extracted from exact locations selected from within the continental USA.

\subsubsection{Covid-19 cases dataset.}
 The COVID-19 dataset consisted of daily confirmed cases and daily death tolls both at the state and at the national levels. These numbers were collected from the CDC (Center for Disease Control and Prevention) website \cite{CDC2020}, and from the COVID Tracking Project initiated at the Atlantic Monthly Group \cite{DataDown37:online}, from March 19, 2020, to August 31, 2020.

\subsection{Preprocessing}
 After extracting the USA tweets, the data were preprocessed. The preprocessing in this study involved converting text to lowercase, removal of punctuation, stop words, numeric values, and ideograms and links. Removing stop words, ideograms, punctuation and numeric values can improve the performance of sentiment analysis techniques \cite{Rustam2021}. Each step is briefly explained below.
\subsubsection{Convert to lowercase.}
Converting all data to lowercase helps in the preprocessing and in later stages of natural language processing when parsing through the data. In this study, we used the \texttt{lower} function part of the Python regular expression module. For example, when applying the function, the words ``COVID", ``Covid" are all converted to ``covid".
\subsubsection{Remove stop words.}
 In natural language processing, stop words are words that if removed do not change the context of a sentence, as for example the words ``the", ``a", ``an" and ``in". Here we used the Natural Language Toolkit (NLTK) library stop word corpus. \cite{Bird2009NLTKbook,Perkins2014Python}.
\subsubsection{Remove punctuation.}
Punctuation characters do not contribute to the sentiment analysis. We removed punctuation, as for example
! " \# \$ \% \& ' (  * + , - . / : ; $<$ = ? @  [  \{ $|$  
from the COVID-19 tweets.
\subsubsection{Remove numeric values.}
Numeric values in the tweets do not contribute to the sentiment embedded in the text. Therefore we removed all numeric values such as 12345, which are not valuable for text analysis.
\subsubsection{Remove ideograms and links.}
We also removed ideograms such as smile faces, flags, etc. Additionally, links in tweets do not contribute to sentiment analysis, hence they were also removed. \\

Three samples of tweets before and after preprocessing are shown in Table\ref{table-SampleCleanned}. All tweets were subject to preprocessing and the cleaned data were then used for calculating the sentiment score of each tweet.

\begin{table}[h]
{\scriptsize
\caption{COVID-19 tweet samples before and after applying preprocessing steps.}
\label{table-SampleCleanned}
\begin{tabular}{p{9.cm}|p{6cm}}
\hline 
{\bf Original tweets} & {\bf Preprocessed tweets}\\
\hline 
``Last Month, was the most difficult time in life.  Loosing my Grandmother was harder than recovering from Covid.  Tomorrow will make a month, since she has gone on to be with the Lord. Https://t.co/eZAmYDvyj7" & "last month difficult time life loosing grandmother harder recovering covid tomorrow make month since gone lord"  \hspace {0.2cm}\\
``To help protect residents from the coronavirus, Volusia County Government will distribute 119,000 surgical face masks beginning Thursday, July 9. Residents may stop by these locations and pick up two masks per https://t.co/tcHUAsvgle" & 

``help protect residents coronavirus volusia county government distribute surgical masks beginning thursday july residents may stop locations pick two masks"
\\
``Thank you to Finley Creek Vineyards, Zionsville Chamber and Westfield Chamber for hosting a wonderful and much-needed in-person luncheon today.  Social-distancing was practiced and a lot of smiles were shared behind https://t.co/HfLRj5SdbI" &
``thank finley creek vineyards zionsville chamber westfield chamber hosting wonderful much needed in person luncheon today social distancing practiced lot smiles shared behind"
    \\
\hline 
\end{tabular}
}
\end{table}

\subsection{Word cloud} 
Word cloud is a technique for visualizing the level of prominence of frequent words in a text, with large font sizes used for more frequent words. In addition to the previous preprocessing, we also applied stemming and tokenization to the dataset. The word cloud we obtained from the tweets data by applying the Python \texttt{WordCloud} library. 

\subsection{Sentiment Analysis Methods}

Sentiment analysis is a field of study that uses natural language processing techniques to extract opinions and feelings from written texts \cite{Agarwal2011}. 
We obtain preliminary sentiment analysis results using three different lexicon-based methods: TextBlob, AFINN, and SentimentR. A brief description of each method is as follows:

\subsubsection{TextBlob.} TextBlob is an open-source Python library for performing various natural language processing (NLP) tasks on textual data \cite{Loria2020textblob}. It provides a simple Application Programming Interface (API) for performing common NLP tasks such as part-of-speech tagging, noun phrase extraction, sentiment analysis, classification, and translation. 
For sentiment analysis, TextBlob uses Pattern library and NTKL toolkits. The sentiment dictionary of TextBlob consists of 2,918 words annotated with polarity, subjectivity and intensity scores.
TextBlob determines the polarity (positivity or negativity) of a text along with its subjectivity. A sentiment score between 1 and -1, defined as {\sl polarity}, is assigned to the text depending on the most commonly occurring positive (good, best, excellent, etc.) and negative (bad, awful, pathetic, etc.) adjectives. In addition to the sentiment score, {\sl subjectivity} is also determined. Subjectivity quantifies the amount of personal opinion and factual information contained in the text. The subjectivity value can be a number between 0 and 1. A higher subjectivity means that the text contains more personal opinion and less factual information, and a low subjectivity value means less personal and more factual information. 

\subsubsection{AFINN.}
Afinn is a lexicon-based sentiment analysis approach developed by Finn A. Nielsen \cite{Nielsen2011afinn}. It contains more than 2477 words with a valence (polarity) associated with each word. The words in AFINN's lexicon are scored for valence within the range from -5 (very negative) to 5 (very positive), where a positive score indicates positive sentiment and a negative score indicates negative sentiment. For example, the sentence ``Face covering is good and bad" will result in a score of 0 (neutral sentiment) and the sentence ``Face covering is terrible and bad" will result in a score of -6 (negative sentiment), and the sentence ``Face covering is good and beautiful" will result in a score of 6 (positive sentiment).

\subsubsection{SentimentR.} SentimentR is also a lexicon-based sentiment analysis approach developed by Tyler Rinker \cite{Rinker2019SentimentR}. It is a dictionary lookup approach that tries to incorporate weighting for valence shifters (i.e., negators, amplifiers (intensifiers), de-amplifiers (downtoners), and adversative conjunctions). The lexicon contains 11,709 words, whose individual scores may take values between -2 and 1. 
The authors in Ref.~\cite{Ikoro2018SentimentRApp} used SentimentR to analyze sentiments expressed by energy consumers on Twitter.\\

Our preliminary results indicate that the sentiment scores of the three methods are comparable with each other within their different ranges. These results are summarized in Table \ref{tableTweetsSample} where the scores from the three methods are shown for a sample of six tweets. Additional results comparing the three methods are shown in Table~\ref{tableTweetCount}, displaying the number of tweets classified as positive, neutral and negative for the total number of tweets in the dataset. Overall the three approaches yielded a similar trend in the number of positive, neutral and negative tweets. SentimentR tended to extract more negative sentiments than TextBlob and AFINN because it has a large number of negative words in its dictionary. From this point on we will focus on the TextBlob sentiment lexicon results.

\begin{table}[]
\begin{center}
\scriptsize
\caption{Sample of tweets with sentiment scores from TextBlob, AFINN and SentimentR}
\label{tableTweetsSample}
\begin{tabular}{p{8cm}p{1.6cm}p{1.3cm}p{1.3cm}p{1.3cm} }
\hline
Sample Tweets & TextBlob Subjectivity & TextBlob Polarity & AFINN Polarity & SentimentR Polarity \\
\hline
``Thank you to Finley Creek Vineyards, Zionsville Chamber and Westfield Chamber for hosting a wonderful and much-needed in-person luncheon today.  Social distancing was practiced and a lot of smiles were shared behind. https://t.co/HfLRj5SdbI " & ~~1.00 & 1.00 ~~~ Positive & 9.00 ~~~ Positive  & 0.794 ~~~ Positive \vspace{1.4cm} \\
``We would like to take the time to thank @kimberlasskick for our wonderful Naomi mask…we got them today. Now we will be well protected from the craziness known as the COVID. \#GlowedUp \#WeLoveNaomi" & ~~1.00 & 1.00 ~~~ Positive & 9.00 ~~~ Positive & 0.637 ~~~ Positive \vspace{1.3cm} \\
``To help protect residents from the coronavirus, Volusia County Government will distribute 119,000 surgical face masks beginning Thursday, July 9. Residents may stop by these locations and pick up two masks per https://t.co/tcHUAsvgle" & ~~0.00 & 0.00 ~~~ Neutral & 2.00  ~~~ Positive  & -0.033 ~~~ Negative \vspace{1.3cm} \\
``How Colleges Should Help Close the Digital Divide in the COVID-Era https://t.co/0TndpE9BCL via @latenightparent" & ~~0.00 & 0.00 ~~~ Neutral & 2.00 ~~~ Positive & 0.0 ~~~ Neutral  \vspace{1.cm} \\
``Last Month, was the most difficult time in life.  Loosing my Grandmother was harder than recovering from Covid.  Tomorrow will make a month, since she has gone on to be with the Lord. Https://t.co/eZAmYDvyj7" & ~~0.392 & -0.025 ~~~ Negative & -1.00 ~~~ Negative & -0.025 ~~~ Negative \vspace{1.3cm} \\
``States, red and blue, are manipulating statistics in order to open early, inflating the number of tests, underreporting the number of cases and deaths. Virus doesn't care.  Virus doesn’t give a shit. https://t.co/a3dZitWf0W https://t.co/W8X955oYVC" & ~~ 0.340 & -0.020 ~~~ Negative & -7.0 ~~~ Negative& -0.323 ~~~ Negative \\
\hline
\end{tabular}
\end{center}
\end{table}

\begin{table}[h]
\small
\caption{Comparative table showing the number of tweets labelled as positive, neutral and negative by TextBlob, AFINN and SentimentR.}
\label{tableTweetCount}
\begin{tabular}{llll}
\hline
Sentiment & TextBlob & AFINN & SentimentR \\ 
\hline
 \# of positive & 47,946 & 41,888 & 53,847\\ 
 \# of neutral   &  22,520 &  27,406 & 9,525\\ 
 \# of negative  &  14,619   & 15,793   & 21,715\\ \hline
\textbf{Total}      & 85,085  & 85,085 & 85,085  \\ 
\hline
\end{tabular}
\end{table}

\subsection{Correlation Analysis}
To investigate the relationship between the tweets' sentiment scores and both the daily new COVID-19 cases and death toll, we apply the Pearson correlation test, which provides a quantitative measure of the linear dependency between two variables.  \cite{follmann2018multimodal,Scipy:online} . It is given by 

\begin{equation}
    r = \frac{\sum (x-\bar{x})(y-\bar{y})}{\sqrt{\sum (x-\bar{x})^2 \sum (y-\bar{y})^2}} 
\end{equation}
where $x$ and $y$ are the sample vectors, $\bar{x}$ and $\bar{y}$ are the corresponding means. Additionally, we used the two tail t-test to assess the significance of the variables.

\section{Results}
\subsection{Tweets Distribution and Sentiment Scores}
\begin{figure}[h]   
\centering
\includegraphics[width=.6\textwidth]{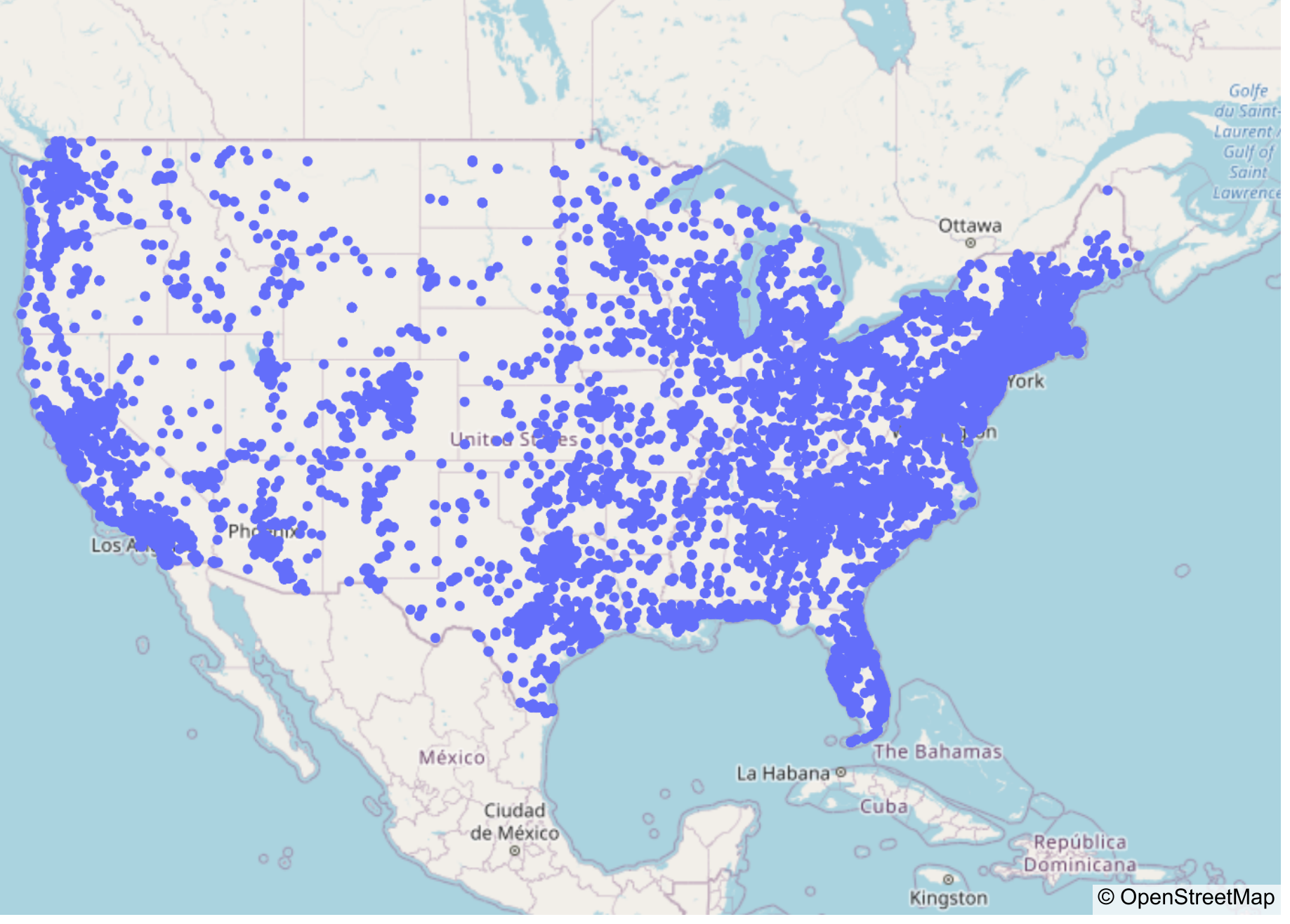}
\caption{Location of COVID-19 related tweets with geo-location available in the USA (blue dots).} 
\label{figUSmap}
\end{figure}

The texts we analyzed were obtained from tweets originated from locations distributed across the USA, as illustrated in Figure~\ref{figUSmap}. The number of tweets per region is directly related to the size of the population of each region. Highly populated communities such as the East and West Coast yield denser distribution and less populated places such as states located in the north-central region yield less dense distributions. 
The top four most populous states (California, Texas, Florida, and New York) account for almost 50\% of the COVID-19 related tweets. California and New York account for 17\% and 16.4\%, respectively, while Florida and Texas account for 5.3\% and 7.3\%, respectively. 

 In terms of the lexicon used in the COVID-19 related tweets, those with high polarity scores tend to use positive words (mostly adjectives) such as ``greatest,” ``best,” ``grateful,” ``perfect,” and ``wonderful.” In contrast, tweets with low polarity scores tend to use negative words (also mostly adjectives) such as ``worst,” ``terrible,” ``killed,” ``no time to waste.” The sentiment scores of polarity ranging between -1 and 1 across our dataset were classified as negative (sentiment score $<$ 0), neutral (sentiment score $=$ 0) or positive (sentiment score $>$ 0).  
  Table \ref{tableStateTweetCount} shows the numbers for tweets classified as positive, negative or neutral for the USA and the top ten populous states. 
  
\begin{figure}[h]   
\centering
\includegraphics[width=0.75\textwidth]{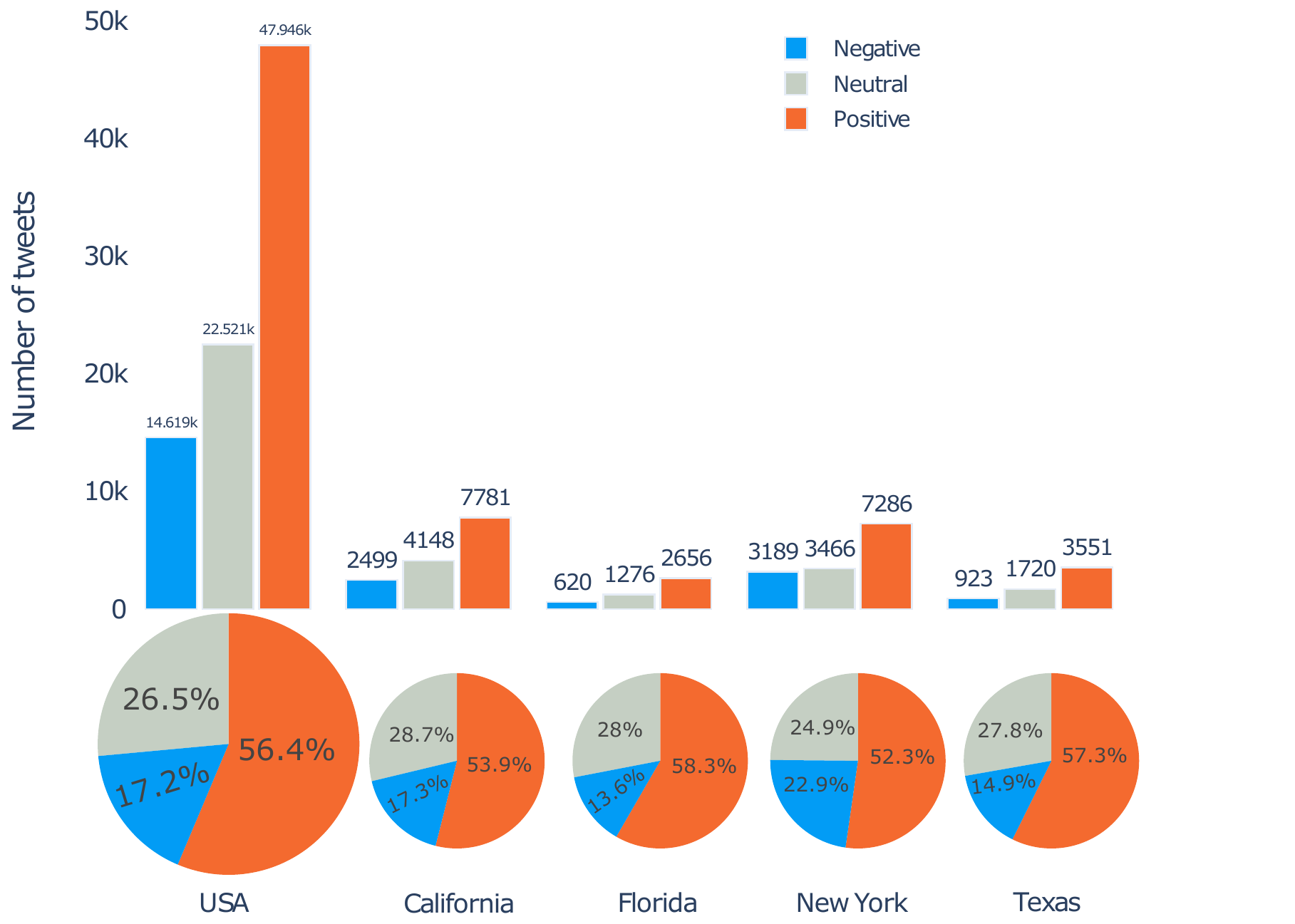}
\caption{Distributions of positive, negative, and neutral tweets in the USA and in the States of California, Florida, Texas, and New York.}
\label{figStatesTweets}
\end{figure}

For the purpose of graphical visualization we show in Figure~\ref{figStatesTweets} that out of the 85,085 tweets in the USA dataset, 56.4\% expressed a positive sentiment, 26.5\% expressed a neutral sentiment, and 17.2\% expressed a negative sentiment. Different but similar distributions were observed at the state level as indicated by the bar and pie graphs for the top four states. These results suggest that humans tend to express positivity, even during difficult times as is the case in a pandemic.

\begin{table}[t]
\small
\sf\centering
\caption{Top ten US states tweets.}
\label{tableStateTweetCount}
\begin{tabular}{lllll}
\hline
 & Total \# of tweets & \# of positive & \# of neutral & \# of negative  \\ 
\hline
USA & 85,085 &47,946 & 22,520  & 14,619   \\
\hline
California &14,428 &7,781 & 4,148 & 2,499   \\
Texas &6,194 &3,551 &1,720 & 923 \\
Florida &4,552 & 2,656&1,276 & 620 \\
New York & 13,941& 7,286& 3,466 &3,189 \\
Pennsylvania & 2,264& 1,275& 622&367 \\
Illinois & 2,356& 1,322& 639& 395 \\
Ohio & 1,657& 985& 433&239 \\
Georgia & 3,329& 1,889& 952&494 \\
North Carolina & 1,830& 1,103&488 &239 \\
Michigan & 1,188 & 691&166 &331 \\
\hline
\end{tabular}
\end{table}

\begin{figure}[h]   
\centering
\includegraphics[width=.55\textwidth]{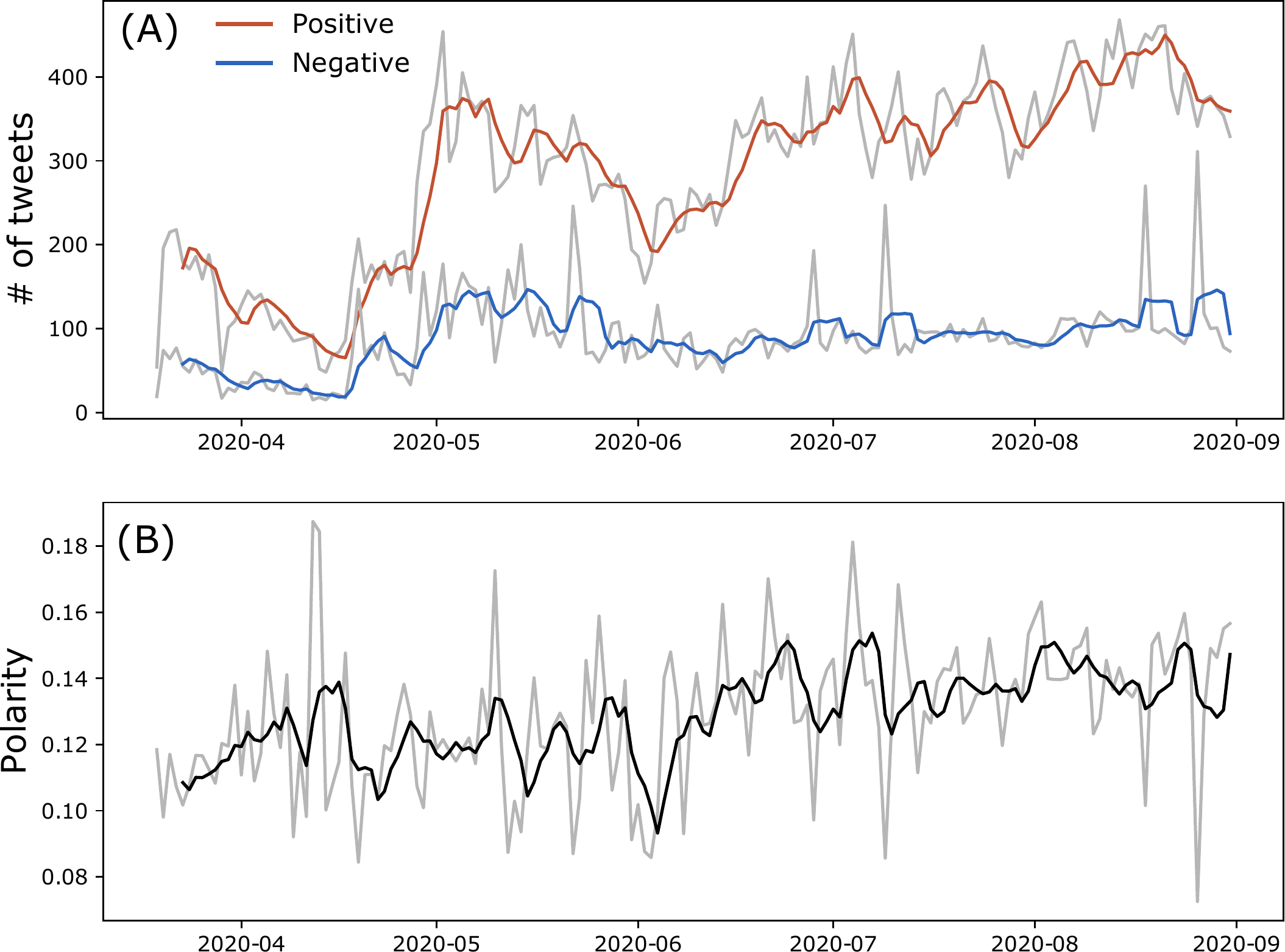}
\caption{(A) Five-day moving averages of the number of positive and negative tweets over time. (B) Average daily polarity. Grey lines represent daily values.} 
\label{figNumbTweetsPolarity}
\end{figure}

In order to provide an overall view of the effects of the pandemic in connection with the number of tweets on a day-to-day basis, we show in Figure~\ref{figNumbTweetsPolarity} a five-day moving average of the number of positive and negative tweets over time (top graph). 
Here we notice that the daily number of positive tweets consistently surpasses the number of negative ones.  
The bottom graph in the same Figure~\ref{figNumbTweetsPolarity} displays the five-day moving average for the daily polarity score, showing that the daily average polarity remains positive over time. 
%

\begin{figure}[h]   
\centering
\includegraphics[width=.65\textwidth]{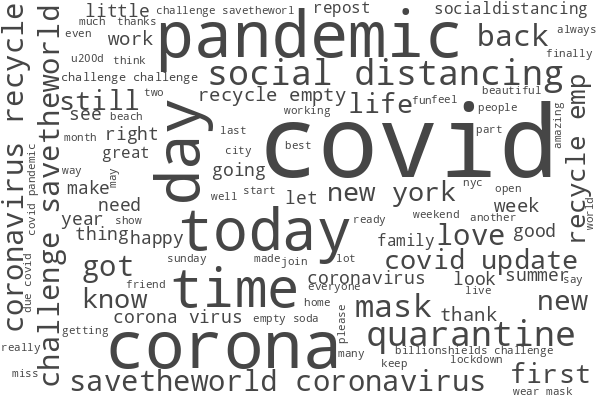}
\caption{Word cloud of USA tweets.} 
\label{figWordCloud}
\end{figure}

The word cloud for the dataset is shown in Figure~\ref{figWordCloud}. In this figure, the higher the presence of a word in the tweets, the larger the font used to display it in the cloud, providing a visual account of the most-used vocabulary in the day-to-day challenges posed by the coronavirus. Not surprisingly, the word ``covid" is the most directly related to COVID-19, followed by ``pandemic," ``corona," ``today," ``social distancing," ``quarantine," etc. But we can also see positive words such as ``love", ``family" and ``beautiful," for example.

\subsection{Subjectivity and Polarity reveal opinion preference}
While polarity gave us an account of how positive or negative peoples' sentiments were, we also looked at the subjectivity score to check whether people expressed factual information (low score) or opinion (higher score) in their tweet messages. Low subjectivity score tweets tend to use more factual vocabulary compared to those with higher polarity scores.
Samples of subjectivity and polarity scores tweets extracted from the dataset can be found in Table \ref{tableTweetsSample}.

\begin{figure}[h]   
\centering
\includegraphics[width=0.65\textwidth]{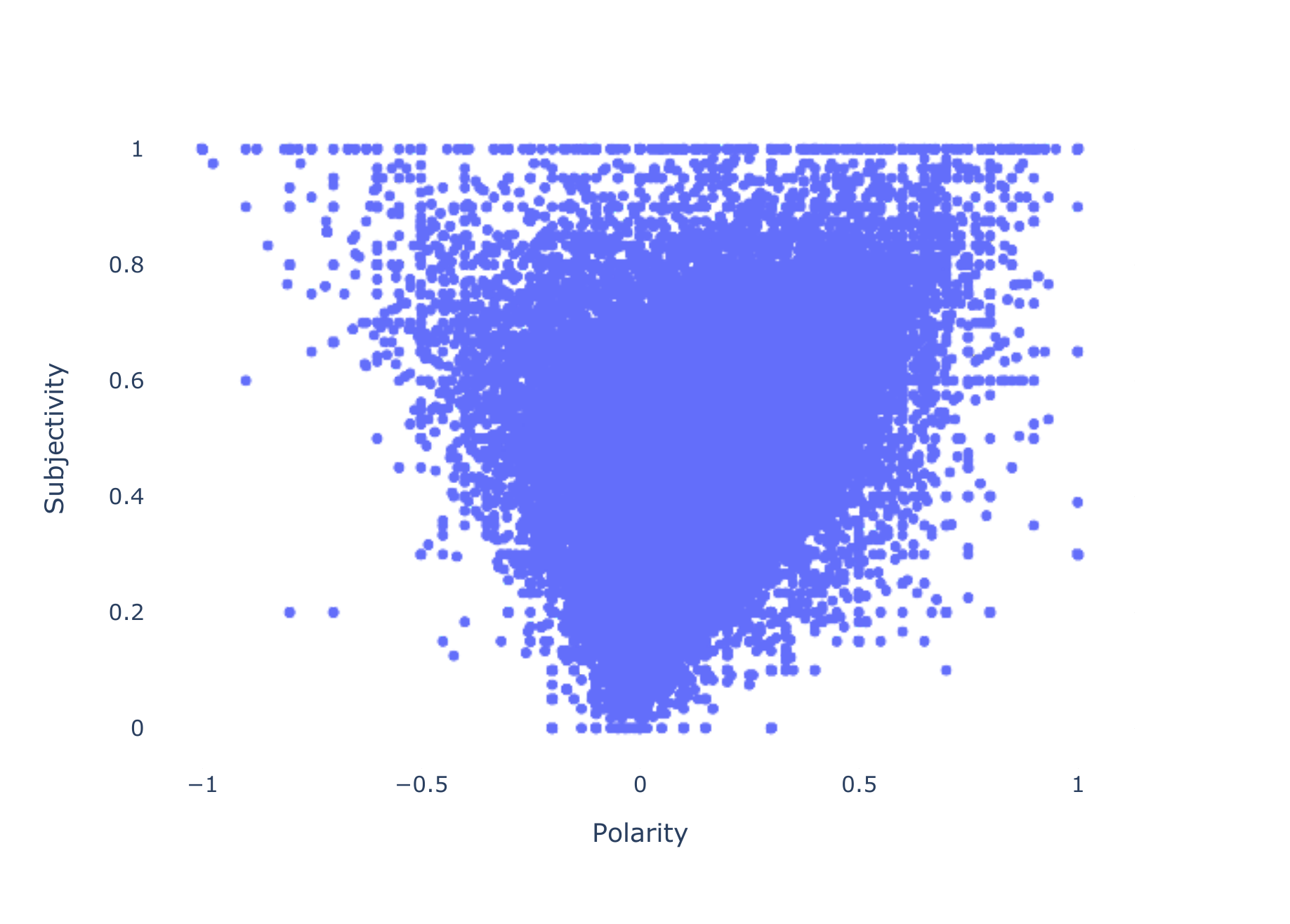}
\caption{USA polarity vs. subjectivity scores illustrates a skewed distribution with more points towards the positive polarity scores and higher subjectivity scores.} 
\label{figUSscatter}
\end{figure}

 A graphical representation of how polarity and subjectivity are related to each other is displayed in the scatter plot of Figure~\ref{figUSscatter}. We observe a skewed distribution with more points towards the positive polarity scores and higher subjectivity scores ($>0.5$), suggesting that the more positive-oriented a tweet is, the more opinion-oriented its meaning will be. 
This can be quantified by measuring the proportion of tweets with positive polarity and subjectivity greater than 0.5 (Table~\ref{table2}). Among the tweets with subjectivity higher than 0.5, positive tweets comprise 78.3\% accounting for 24\% of the total number of tweets. The combination of neutral and negative tweets account for 21.7\% of the tweets with subjectivity above 0.5. This corresponds to 6.7\% of the total number of tweets. These numbers indicate that nationwide, people are more likely to express their positive rather than their negative opinions. 

\begin{table}[h]
\sf\centering
\caption{Number of tweets with  subjectivity higher than $0.5$.}
\label{table2}
\begin{tabular}{llll}
\toprule
Subjectivity \textgreater 0.5 & Tweet counts & Percent \\ 
\midrule
Polarity \textgreater 0 & 20,524 & 78.3\% \\ 
 Polarity \textless 0    & 4,769  & 18.2\% \\ 
 Polarity = 0             & 922    & 3.5\%  \\ 
\bottomrule
\end{tabular}
\end{table}

\begin{figure}[h]   
\centering
\includegraphics[width=0.8\textwidth]{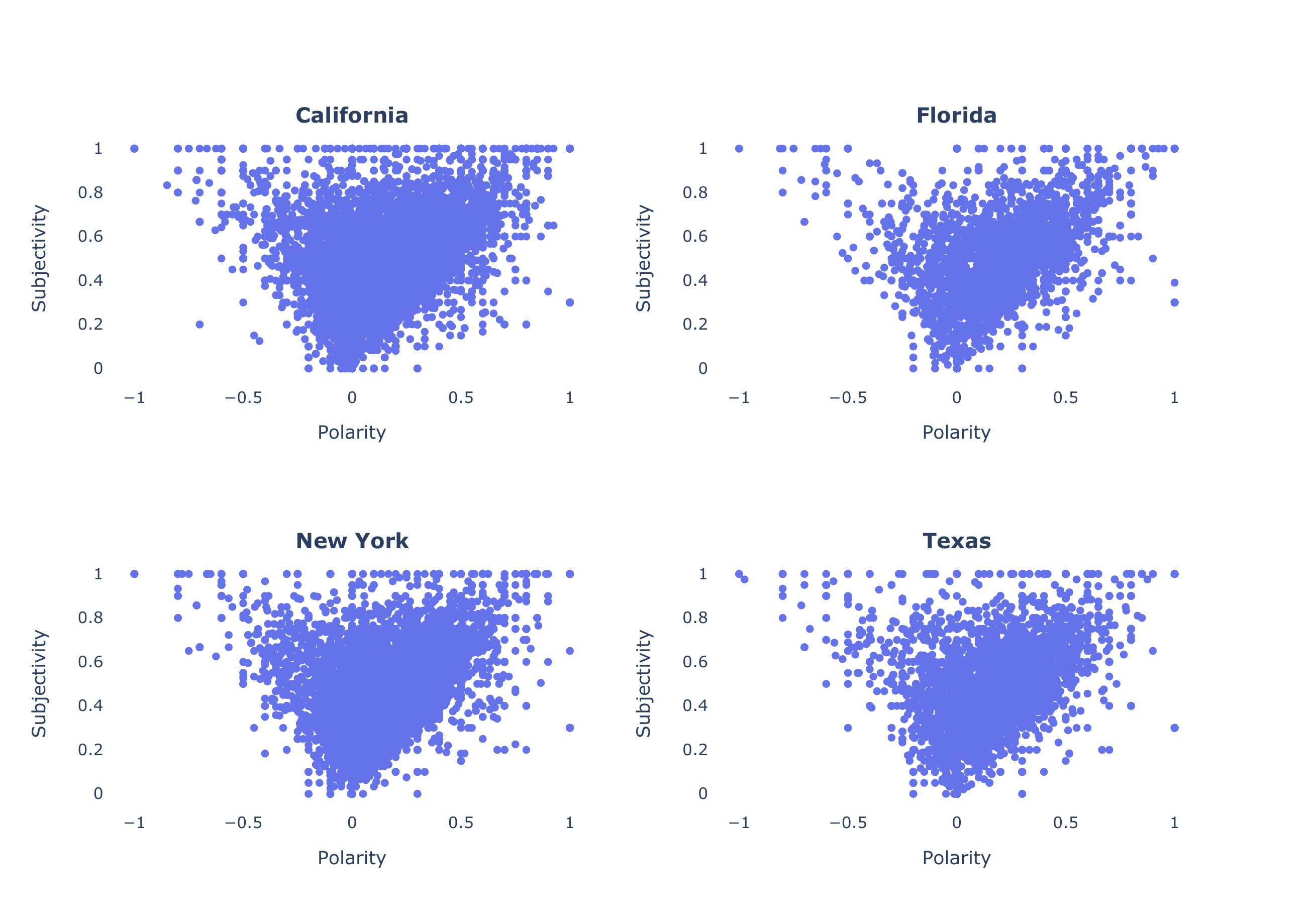}
\caption{Polarity vs. subjectivity scores for California, Florida, New York, and Texas.} 
\label{figScatterStates}
\end{figure}

A similar result is obtained at the State level as displayed in  Figure~\ref{figScatterStates}, which shows the relationship between the subjectivity and polarity scores in California, Florida, New York, and Texas. 
Interestingly, while displaying a tendency of being more positive even under difficult conditions, humans tend to be more responsive to negative than to positive news \cite{Soroka2019}.  

\subsection{COVID-19 cases and sentiment scores in the USA}
It is unquestionable that the COVID-19 pandemic is affecting all aspects of our day-to-day lives, including our emotions. In this section, we present results obtained from using sentiment analysis to investigate the extent of the influence that daily new COVID-19 cases and death tolls are having on our emotions.

\subsubsection{Correlation results.} 
\begin{figure}[h]   
\centering
\includegraphics[width=0.7\textwidth]{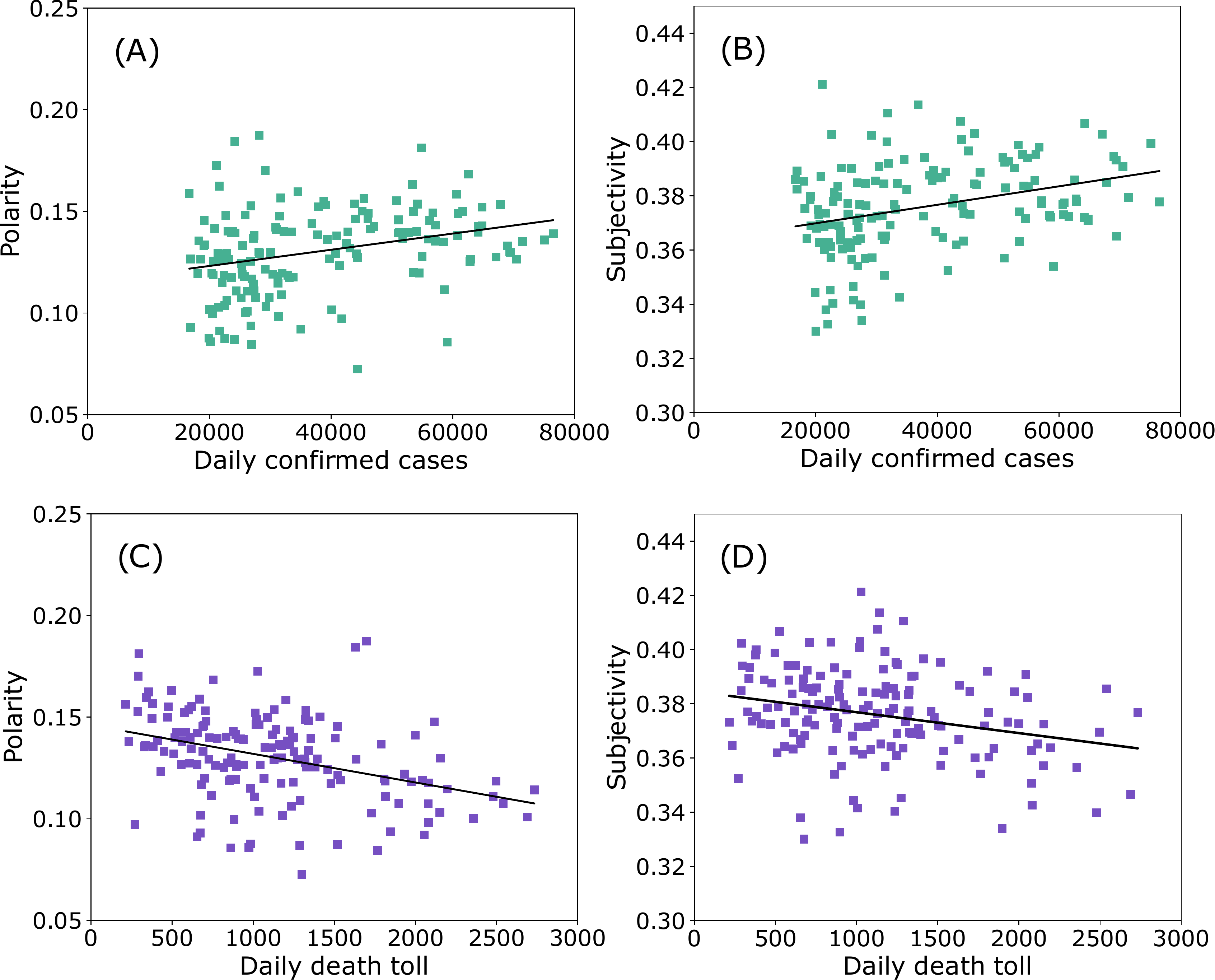}
\caption{Sentiment scores polarity and subjectivity in graphs (A) and (B) {\it vs.} daily confirmed cases, and in graphs (C) and (D)   {\it vs.} daily death toll. (A)-(B): Pearson correlation coefficient r = 0.29, p $<$ 0.0001 and r = 0.32, p $<$ 0.0001, respectively;). 
(C)-(D): Pearson correlation coefficient r = -0.37, p $<$ 0.0001 and r = -0.26, p $<$ 0.001, respectively. From April to August n = 153.} 
\label{figCorr}
\end{figure}

The graphs in Figure~\ref{figCorr} show how COVID-19 daily confirmed cases and daily death toll, relate to polarity and subjectivity scores. They indicate a weak correlation between daily confirmed cases and both polarity (graph~\ref{figCorr}(A)) and subjectivity (graph~\ref{figCorr}(B). These numbers suggest that people's sentiment is affected, but not strongly, by the daily increase of COVID-19 confirmed cases.
Graphs~\ref{figCorr}(C) and~\ref{figCorr}(D) also show a weak negative correlation, but between daily death toll and polarity and subjectivity, respectively. These results, similarly to cases (A) and (B) above, suggest that people's sentiment is also affected, but not strongly, by the increase in the daily death toll due to COVID-19. 
However, a comparison between statistical coefficients of the combined polarity and subjectivity with respect to daily confirmed cases (graphs ~\ref{figCorr}~(A) and (B)), and the combined polarity and subjectivity with respect to daily death toll (graphs ~\ref{figCorr}~(C) and (D)), indicates that the daily death toll has a larger effect on peoples sentiment than new daily confirmed cases. This result can be understood from the perspective that, an increase in the number of deaths poses a more threatening challenge compared with an increase in the number of new cases which, even though threatening, still has the door open for a possible recovery. 

\subsubsection{Temporal results.}
\begin{figure}[t]   
\centering
\includegraphics[width=0.7\textwidth]{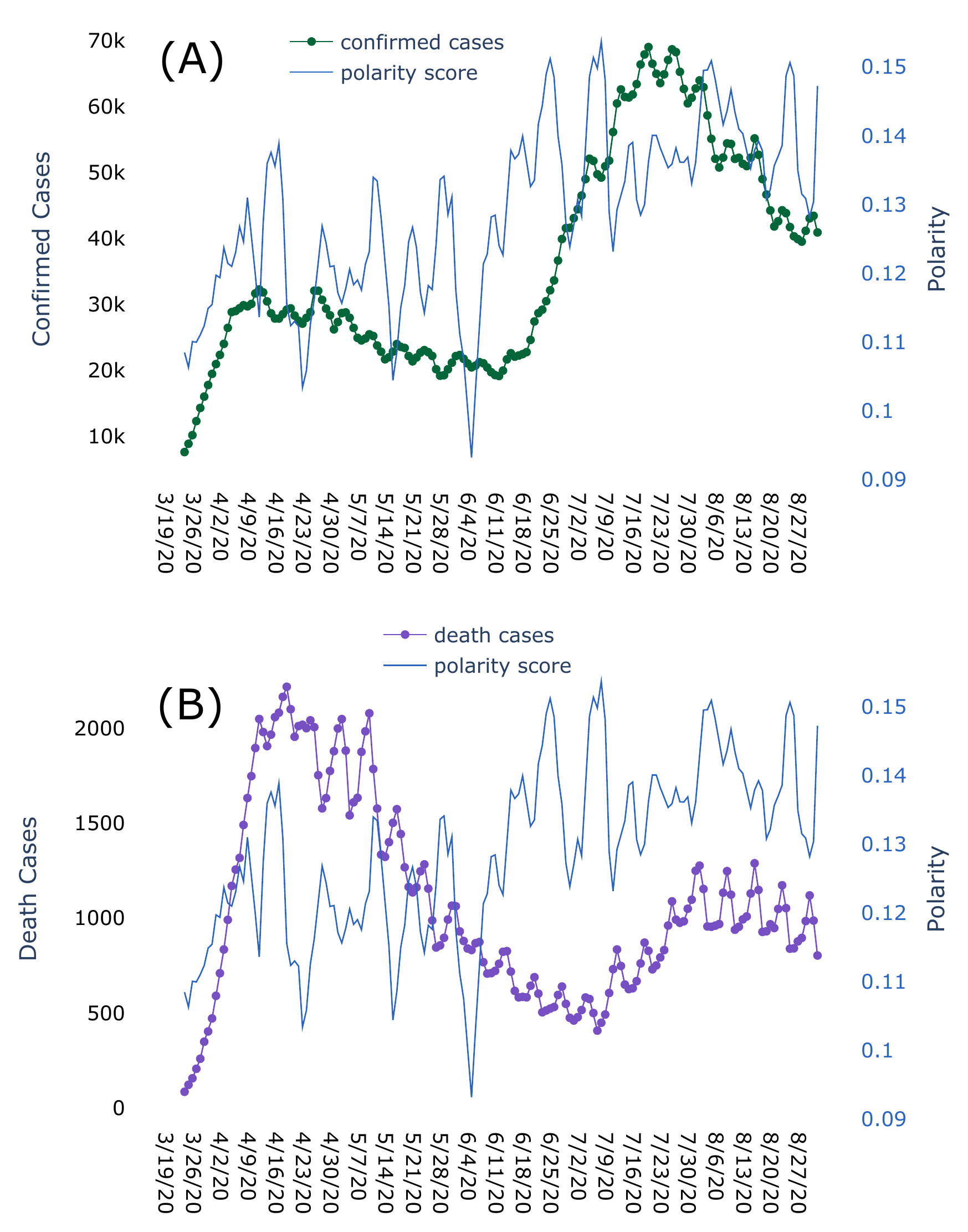}
\caption{Five-day moving average for COVID-19 cases, death tool and polarity score in the USA. (A) Confirmed cases, (B) Death toll. } 
\label{figPolarityCovidCasesDeath}
\end{figure}

We now analyze the time evolution of the polarity score in connection with the number of confirmed cases and the death toll using a five-day moving average from March 19 to August 30, as shown in Figure~\ref{figPolarityCovidCasesDeath}. In graph A, the confirmed cases, scaled on the left-hand side y-axis, were at a downtrend fluctuating around 20,000-30,000 from early April to the third week of June. A sharp increase is noticed starting on the third week of June, extending to the end of July, and going down in early August. The polarity, scaled on the right-hand side y-axis, displays a wide range of fluctuations with a clear increase in mid-June. In this case, both the number of confirmed cases and polarity show a good agreement in their overall trends.
This is consistent with the weak, but significant positive correlation between the confirmed cases and the polarity score previously described and shown in Figure~\ref{figCorr}~(A).

As for the number of deaths and polarity, their trends are reversed with respect to each other, as displayed in graph ~\ref{figPolarityCovidCasesDeath}(B). The higher values for the number of deaths happen to be in between 4/1/20 and about 6/3/20, with lower values on the average for the remaining time. For the polarity, however, the lower values happen around the first half of the period and higher values on the second half of the period. This is an indication that when the number of deaths increases, Twitter users are more likely to express mournful feelings, leading to a low polarity score. On the other hand, when the death toll is lower, people seem to be more optimistic so that more positive tweets are generated, leading to a higher polarity score.

\begin{figure}[t]
\centering
\includegraphics[width=0.8\textwidth]{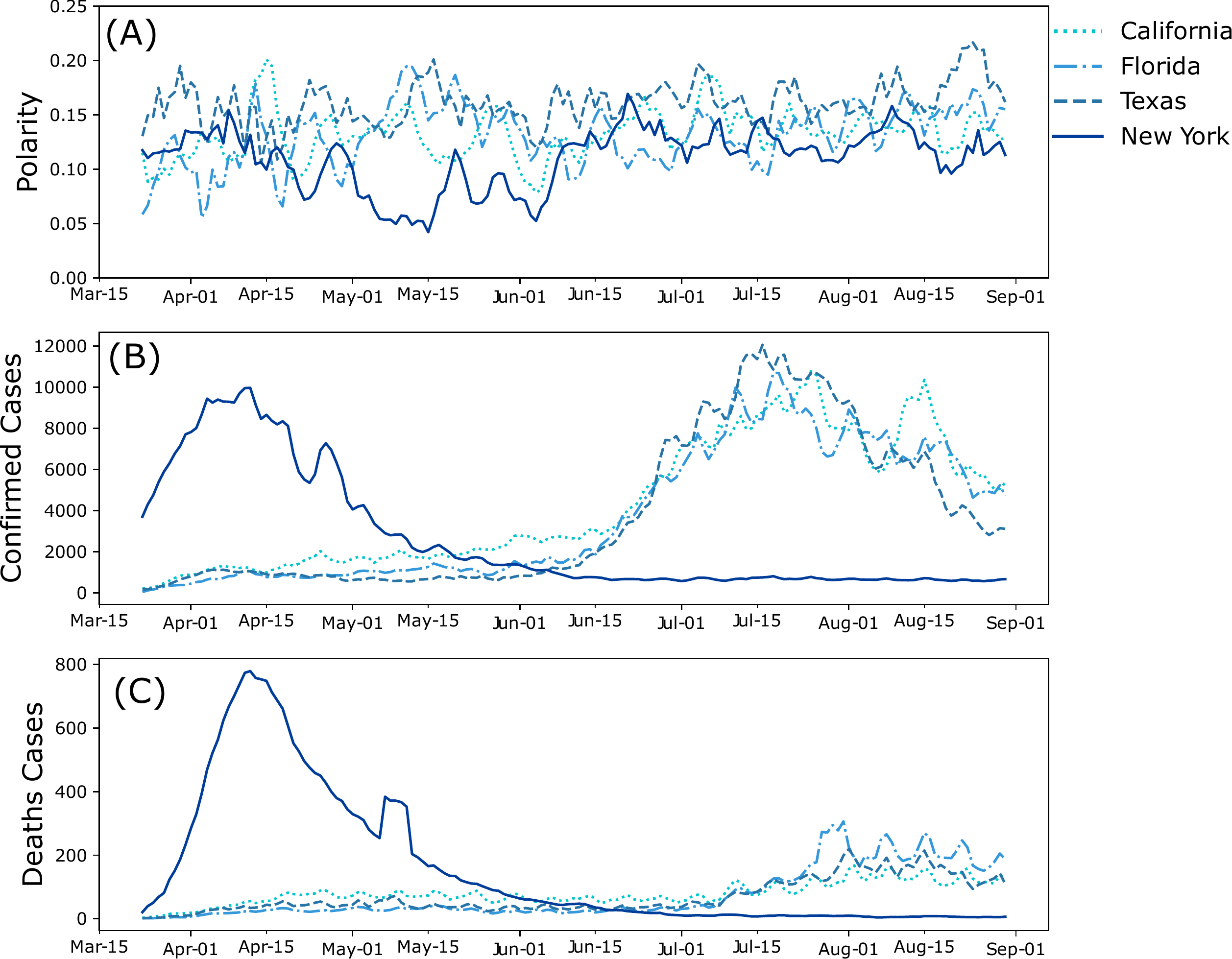}
\caption{Five-day moving average for COVID-19 cases, death tool and polarity score at the state level. (A) Polarity, (B) Confirmed cases, (C) Death toll. } 
\label{fig-StatePolarity_MA5}
\end{figure}

Aiming at providing information about the sentiments of people in largely affected areas, we now analyze the number of COVID-19 cases, the death toll and the polarity scores for the top four most populous states. Figure~\ref{fig-StatePolarity_MA5}(A) presents the evolution of the polarity scores from March 19 to August 30, 2020, for California, Florida, Texas and New York, with estimates of daily polarity fluctuations between 0.05 and 0.23. In graph (B), the confirmed cases show a distinct evolution for New York compared to the other three states showing a peak of confirmed cases in mid-April, while the other three states show a peak in mid-July, consistently with the fact that New York was affected first by the pandemic. Graph (C) shows a similar temporal evolution for the daily number of deaths with New York again exhibiting a peak in mid-April followed by a downward trend for the rest of the period. Interestingly, the other three states had a peak in the number of death cases in early August but not as prominent as was the case in New York. This is probably a consequence of the fact that by August health professionals and hospitals were better prepared and equipped to treat COVID-19 patients.  

\subsubsection{Polarity score and significant events.}
\begin{figure}[t]
\centering
\includegraphics[width=0.9\textwidth]{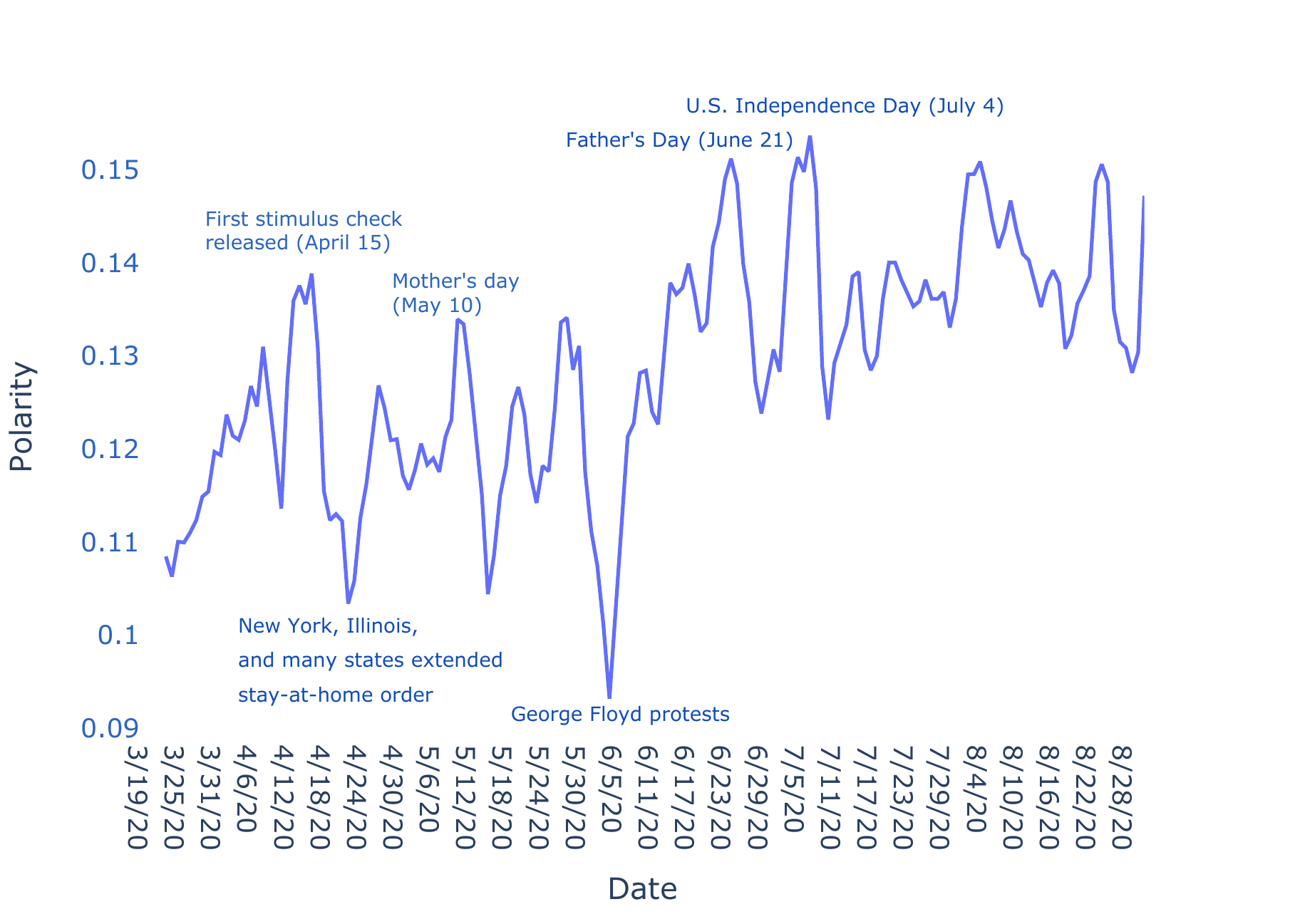}
\caption{Five-day moving average of the polarity score and significant events.} 
\label{polarity_MA5}
\end{figure}

While the predominant sentiment analysis factor determining the oscillations in polarity show a direct connection with the pandemic itself, other factors connected or not with the pandemic may have a punctual influence on the polarity. Figure~\ref{polarity_MA5} shows the time evolution of the polarity displaying also the dates of two events promoting lower polarity values: 
(i) The extended stay-at-home order implemented in New York and Illinois: On April 16, New York Governor Andrew Cuomo extended the state's stay-at-home order and school closures through May 15 \cite{NewYorkT22:online}. Also, on April 23, Illinois Governor J. B. Pritzker extended the statewide stay-at-home order through May 29 \cite{Pritzker99:online}. These orders were implemented in a number of other states as well. The polarity score decreased from April 17 to April 25, 2020. 
(ii) The George Floyd protests, an event with no known connection with the pandemic: From May 26th to June 10th around half a million people joined protests in 550 places in the USA in connection with the death of George Floyd in Minneapolis on May 25th, 2020 \cite{BlackLiv81:online}.
Additionally, we included two other events that seemingly promoted the increase in polarity values: 
(iii) The issue of the first coronavirus stimulus checks coinciding with the Easter celebration:  When the first coronavirus stimulus checks were deposited on the week of April 13th \cite{Firstcor92:online}, along with the celebration of Easter (April 12nd), we observe an increase in the polarity score (April 12nd to April 16th, 2020).
(iv) The Fourth of July holiday: This is a traditional holiday celebrated with joy in the whole country, and therefore it would be expected to have a noticeable positive effect on the polarity. In fact, the polarity score increased from July 2 to July 10.  Interestingly, significant events such as new regulations from the government (stimulus check or stay at home orders), the celebration of certain holidays and social conflicts that directly affect the sentiment of the public, can be captured through the change of the sentiment score.

\section{Conclusion}
The difficulties we are presently facing with the COVID-19 pandemic are not new. Back in 1918, for example, the human race was plagued by an H1N1 virus pandemic that caused more than 50 million deaths worldwide, of which 675,000 occurred in the United States \cite{H1N1Pandemic:online}.  At the time, with no vaccine, no medication and no infrastructure to alleviate the symptoms, control measures were limited to isolation, quarantine, personal hygiene, disinfectants and restrictions of gathering. These are striking similarities with the COVID-19 pandemic, but there are differences as well, including age groups with higher vulnerability, and easiness of travel which of course plays a role in the spread of the virus. Additionally today we have more effective communication, which can help disseminate useful information with preventive effects, but also damaging information which might cause people to underestimate the risks of the COVID-19 virus. 

Social media has also given rise to plenty of venues for people to express their concerns, fears, joy and happiness in ways not even though possible some 20 years ago. Among others, microblogging is one consisting of shared online posting of short texts containing personal experiences and emotions.  Large amounts of data from microblogging services such as Twitter make them a fascinating source for opinion mining and sentiment analysis. We used TextBlob to calculate each tweet's subjectivity and polarity score and to classify them into positive, negative, and neutral. We presented a comprehensive investigation on the sentiment distribution in the US as a whole and among the four most populous states.
In this work, we also investigated the relationship between the sentiment score and the COVID-19 cases in the United States. Our results indicate that there is a link between sentiment scores and COVID-19 confirmed cases and the death toll in the USA. Significant events, such as new regulations from the government, celebration of important holidays, and social conflicts, can directly affect the public's sentiment. A striking result of this work points to the realization that, even during difficult and challenging times, people tend to express more positive sentiments.



\bibliography{main}

\end{document}